\documentclass[iop]{emulateapj}

\usepackage{amsmath}
\usepackage{amsfonts}
\usepackage{graphicx}
\usepackage{xcolor}
\usepackage{mathtools}
\usepackage{afterpage}
\usepackage{nicefrac}
\usepackage{listings}
\usepackage{wasysym}
\usepackage{stix}
\usepackage{ifsym}
\usepackage{url}

\usepackage{times}

\newcommand{\beq}{\begin{equation}}
\newcommand{\eeq}{\end{equation}}
\newcommand{\bea}{\begin{eqnarray}}
\newcommand{\eea}{\end{eqnarray}}
\newcommand{\mrm}{\mathrm}

\newcommand{\mdot}{\dot{M}}
\newcommand{\msun}{\mathrm{M}_\odot}
\newcommand{\lsun}{\mathrm{L}_\odot}
\newcommand{\rsun}{\mathrm{R}_\odot}
\newcommand{\msunpyr}{$\mathrm{M}_\odot{\rm yr}^{-1}$}

\newcommand{\Rsp}{R_{\rm sp}}

\newcommand{\ohat}{\hat{\mrm{o}}}

\newcommand{\inv}{^{-1}} 

\shortauthors{Ro}

\begin{document}

\title{ The Wolf-Rayet Stellar Response To The Iron Opacity Bump: \\ Envelope Inflation, Winds, and Microturbulence}

\author{Stephen Ro\email{sro@berkeley.edu} }
\affil{Astronomy Department, University of California at Berkeley, Berkeley, CA 94720-3411, USA}

\begin{abstract}
Early-type Wolf-Rayet (WR) stellar models harbour a super-Eddington layer in their outer envelopes due to a prominent iron opacity bump. In the past few decades, one-dimensional hydrostatic and time-steady hydrodynamic models have suggested a variety of WR responses to a super-Eddington force including envelope inflation and optically-thick winds. In this paper, I study these responses using semi-analytical estimates and WR models from both \texttt{MESA} and \cite{rm16}; four conclusions are present. First, early-type WR stars do not harbour inflated envelopes since they have either strong winds or insufficient luminosities. Second, the condition for an opacity bump to harbour a sonic point is expressible as a minimum mass loss rate, $\mdot_{\rm sp}(L_*)$. In agreement with \cite{phdthesis} and \cite{2018A&A...614A..86G}, the majority of galactic early-type WR stars can harbour sonic points at the iron opacity bump. However, about half of those in the Large Magellanic Cloud cannot given typical wind parameters. Third, WR winds driven by the iron opacity bump must have mass loss rates that exceed a global minimum of $10^{-5.8}-10^{-6}$\,$\msun$\,yr$\inv$. Lastly, the observed early-type WR distribution follows a simple mass loss relation derived here if the radiation-to-gas pressure ratio is approximately $p_r/p_g\simeq145$ in the wind; a value consistent with studies by \cite{2017A&A...608A..34G} and \cite{2018ApJ...852..126N}. 
\end{abstract}

\section{Introduction}
\label{sec:introduction}
The inclusion of fine iron M-shell transitions in the Rosseland opacity calculations led to the discovery of a prominent opacity bump by the OPAL and the Opacity Project groups \citep{1992ApJS...79..507R,1994MNRAS.266..805S}. Situated at $T\simeq10^{5.2}$ K, the iron bump resolves many outstanding issues regarding the $\kappa$-mechanism in pulsating stars such as the ``bump and beat'' mass discrepancy in Cepheids \citep{1992ApJ...385..685M}, oscillations in $\beta$ Cephei and slowly pulsating B-type stars \citep{1993MNRAS.262..204D,1997ApJ...483L.123C}. For Wolf-Rayet (WR) stars, the iron bump has been suggested to trigger a variety of responses including envelope inflation \citep{1999PASJ...51..417I,petrovic, 2012A&A...538A..40G}, strange-mode oscillations \citep{1993MNRAS.262L...7G, 1996MNRAS.281..406K, glatzel99,2002MNRAS.337..743G}, turbulence \citep{gras16}, and quasi-steady winds \citep{nl02, 2005A&A...432..633G, 2017A&A...608A..34G, 2018ApJ...852..126N, 2018A&A...614A..86G}. These responses are challenging to test observationally since WR winds are optically thick which renders basic stellar parameters invisible (e.g., mass, radius, rotation).

WR spectral analyses employ \texttt{CMFGEN} \citep{1998ApJ...496..407H} and \texttt{PoWR} codes \citep{1985A&A...148..364H, 2002A&A...387..244G} to study WR interiors. These codes compute radiative transfer calculations in the co-moving frame prescribed by a supersonic $\beta$-law velocity structure $v=v_{\infty}(1-R_{20}/r)^{\beta}$ \citep{1975ApJ...195..157C}. The radiative acceleration is treated self-consistently in \texttt{PoWR} \citep{2015A&A...577A..13S}. These studies find roughly half of stars have `hydrostatic' radii, $R_{20}$, (defined where the Rosseland optical depth is $\tau\sim20$) exceed stellar model predictions, $R_*$, by up to an order of magnitude. Consequently, the `hydrostatic surface' temperatures, $T(R_{20})= (L_*/(4\pi\sigma R_{20}^2))^{1/4}$, and escape speeds, $v_{\rm esc}(R_{20})$, are significantly smaller than expected. This is especially problematic if we assume hydrogen-free stars are on the helium-burning main-sequence since their structures are simple \citep{1989A&A...210...93L, 1992A&A...263..129S}.   

The stellar radius is not the only discrepancy seen in WR stars. WC stars are a WR sub-type with strong carbon spectral line features. Evolutionary models predict WC luminosities exceed $10^{5.4}$\,$\lsun$ \citep{2005A&A...429..581M,2012A&A...542A..29G} despite observations of galactic WC stars with $4.9\lesssim\mathrm{log_{10}}(L_*/\lsun)\lesssim5.4$ \citep{2012A&A...540A.144S,2017MNRAS.470.3970Y}. Stellar models predict the luminosities of early-type WN stars (which have nitrogen-dominant spectral lines) establishes the spectroscopic properties \citep{1989A&A...210...93L}, which is not seen in observations \citep{2006A&A...457.1015H}. While the luminosity discrepancy is significant, this investigation considers only the radius discrepancy problem. 

Sub-surface convection is an inefficient transport mechanism in WR stellar models (e.g., see Eq.\,\ref{eq:max_lconv} and \ref{eq:empiricalmesa}). As a result, the iron bump forces a hydrostatic envelope to develop a rarified layer of near-Eddington gas that is extended by \textit{several} core-radii \citep{1999PASJ...51..417I} and, in some instances, a massive, dense, super-Eddington shell \citep{joss, sanyal15}. A number of authors have proposed `envelope inflation' as a partial resolution to the WR radius discrepancy problem \citep{1992ApJ...394..305K, 1996A&A...309..129S, 1996A&A...315..421H, petrovic,2012A&A...538A..40G}. If realized, this phenomenon would have several consequences for stellar and transient observations. For instance, there would be earlier binary interactions \citep{2016A&A...596A..58K}, redder massive stars \citep{2012A&A...538A..40G, sanyal15,2016MNRAS.459.1505M} and `circumstellar' imprints in supernova lightcurves and spectra \citep{2015A&A...575L..10M, 2018arXiv180102056D}.

An alternative response to a near- or super-Eddington force is an outflow. \cite{nl02} argue the iron bump in WR envelopes can harbour a sonic point. The luminosity-mass relation for early-type WR stars is a good approximation for the radiative luminosity beneath the sonic point at the iron bump (\citealt{1989A&A...210...93L, 1992A&A...263..129S}; RM16) which permits the Eddington limit (or sonic point condition) near the surface to be written in terms of a critical opacity $\kappa_* = 4\pi c G M_*L_*\inv$. \cite{phdthesis} and \cite{2018A&A...614A..86G} realize the critical opacity cannot be reached for arbitrarily small densities. Therefore, a sonic point exists above a minimum density and mass loss rate, $\mdot_{\rm sp}$. In agreement, these authors find galactic early-type WR stars can harbour a sonic point at the iron bump (i.e., $\mdot_{\rm gal}\gtrsim \mdot_{\rm sp}$). Both authors use the approximate sonic point condition from \cite{nl02}, which is that sonic points rest above the iron opacity peak (i.e., $\partial\kappa/\partial T>0$). In this paper, I revisit this calculation of $\mdot_{\rm sp}$ with the sonic point condition from \cite{rm16}.

\cite{petrovic} and \cite{rm16} (hereafter called RM16) question whether envelope inflation, a hydrostatic phenomenon, can manifest in a wind. RM16 found early-type WR stellar models with mass loss rates below $\mdot_b$ have structures that resemble inflated envelopes; although, the regions contain highly supersonic velocity fields. Since the authors consider only continuum-driven winds, additional line effects (e.g., Doppler-shifts and line-deshadowing) are neglected in the outflow acceleration; thus, $\mdot_b$ is likely an overestimate. Wind models with $\mdot>\mdot_b$ have sufficiently large densities and small pressure scale heights such that envelope inflation does not manifest. 

In this paper, I compare the critical mass loss relations, $\mdot_{\rm sp}$ and $\mdot_b$, with observations of early-type WR stars in the Milky Way and Large Magellanic Cloud. I begin with an explanation for why radiative envelopes inflate (\S\ref{sec:bubbles} and \ref{sec:unboundbubble}) followed by a semi-analytic description of their structure (\S\ref{sec:unstableenvelopesolutions}). I review the equations for a continuum-driven wind in \S\ref{sec:structure_equations} along with the conditions for a sonic point. The critical mass loss relations for a sonic point and envelope inflation are derived in \S\ref{sec:minimummdotsp} and \ref{sec:bubblemdot}. Results and discussions follow in \S\ref{sec:results} and \ref{sec:discussion}.

\section{Stellar Responses to an Iron Opacity Bump}
We begin by deriving the conditions for envelope inflation (\S\ref{sec:bubbles}) and winds (\S\ref{sec:structure_equations}). These conditions may suggest there are only two stellar responses to a prominent iron bump; this is not the case. There are four possible responses partitioned by two critical mass loss relations in $(\mdot,L_*)$-space. 

\subsection{Hydrostatic Response}\label{sec:bubbles}
The Rosseland opacity tables show multiple bumps due to the partial ionization of Fe ($T_{\rm Fe}\equiv 10^{5.2}$\,K), He ($10^{4.6-4.8}$\,K), and H ($10^4$\,K). 
An opacity bump increases the local Eddington ratio, $\Gamma_r=\kappa L_r / 4 \pi c G M$, until the onset of convection:
\beq
\Gamma_r \ge \Gamma_c\equiv \frac{8(4-3\beta)(1-\beta)}{8(4-3\beta)-3\beta^2}< 1,
\label{eq:convection}
\eeq
where $\beta=p_g(p_g + p_r)^{-1}$ is the ratio of gas, $p_g$, and radiation pressures, $p_r$ (and not related to the $\beta$-law). Convection reduces the radiative luminosity, $L_r$, until the opacity decreases or gas becomes optically thin.

Hydrogen-free WR stars are expected to be massive (7\,-\,35\,$\msun$), very luminous ($10^{4.8}$\,-\,$10^6\,\lsun$), and compact ($\sim$\,1\,$\rsun$) with radiation-dominated envelopes ($\beta\lesssim0.3$) \citep{1989A&A...210...93L, 1992A&A...263..129S}. As a result, convective instability occurs at high Eddington ratios, $\Gamma_c\gtrsim0.7$. The maximum convective luminosity $L_c\simeq4\pi r^2 v_c U$ from subsonic transport $v_c<c_s\simeq\sqrt{4p_r/(3\rho)}$ of thermal energy (density) $U\simeq3p_r$ is a small fraction of the total WR luminosity,
\beq
\frac{L_c}{L_*}\lesssim 0.04\left(\frac{M_*}{25\msun}\right)^{-0.45} \left(\frac{\rho}{10^{-10}{\rm \, g \, cm^3}} \right)^{-0.5}.
\label{eq:max_lconv}
\eeq
The above relation is evaluated using the empirical WR relations in Appendix\,A. Note that the opacity tables are unavailable for $\rho\lesssim10^{-10.5}$\,g\,cm$^3$ near iron bump temperatures. Wave transport may contribute an additional $L_w\sim 0.5L_c$, although strong shock formation is inevitable at this point \citep{2017ApJ...841....9R}. As a result, WR envelopes are effectively radiative in structure. 

\begin{figure}
    \centering
    \includegraphics[width=\columnwidth]{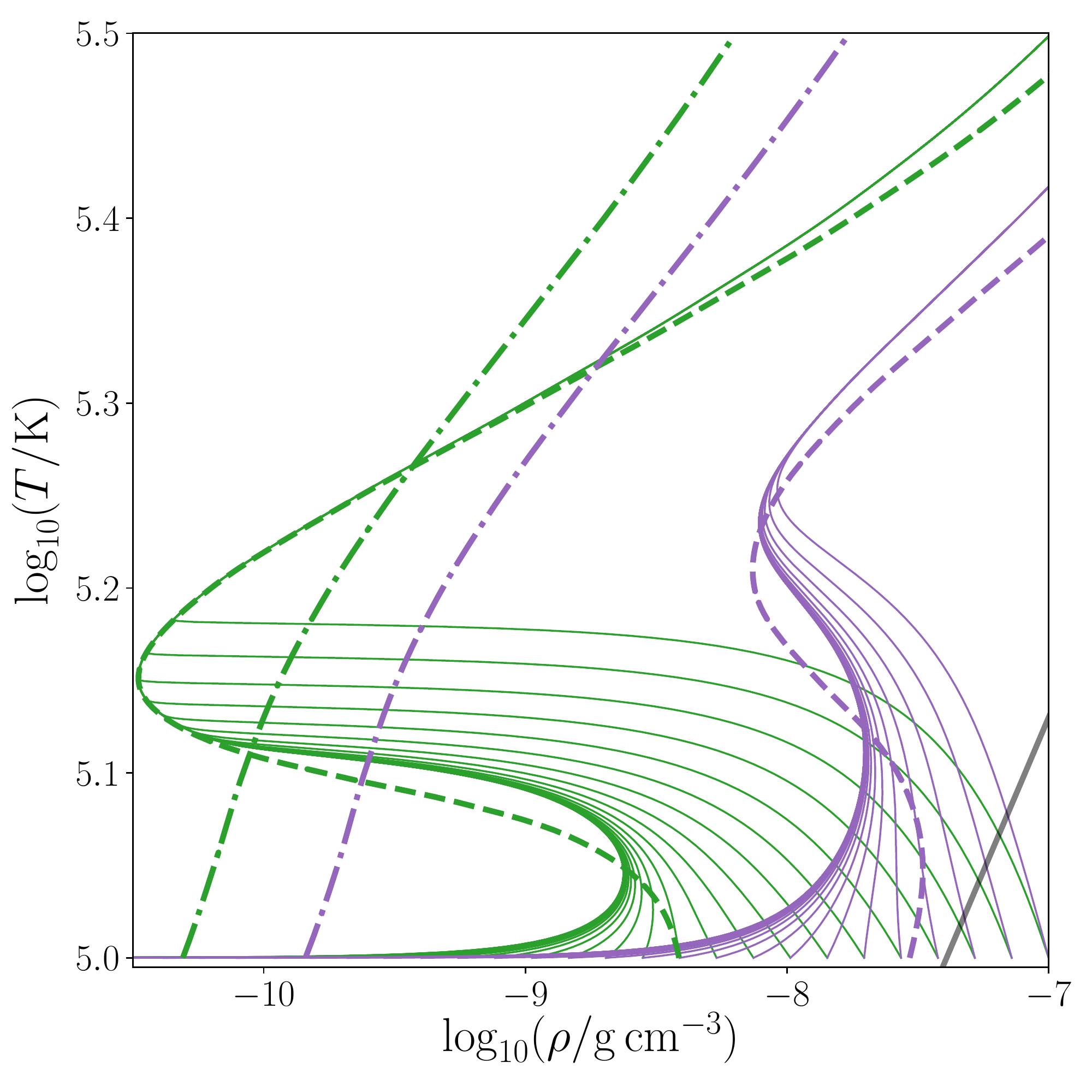}
    \caption{Dashed and solid lines are solutions to Eq.\,(\ref{eq:EddingtonContour}) and (\ref{eq:ode_static}), respectively, for $M_*=10$ (purple) and 25\,$\msun$ (green) pure helium stellar models with solar metallicity. The grey line indicates where gas and radiation pressures are equal. Dot-dashed lines are where the envelope temperature scale height equals the local radius (i.e., $\alpha\equiv H_T/r=1$).}
    \label{fig:rho_T_ode}
\end{figure}

The WR envelope can be well approximated by the equations of continuity, radiative diffusion approximation 
\beq
\frac{1}{\rho}\frac{dp_r}{dr} = -\frac{\Gamma_r v_k^2}{r },
\label{eq:diff}
\eeq
and hydrostatic pressure balance,
\beq
 \frac{1}{\rho}\frac{d p_g}{dr} =  - \frac{v_k^2(1-\Gamma_r)}{r},
\label{eq:hydrostatic}
\eeq
where $p_r=a_r T^4/3$ is the radiation pressure with $a_r$ being the radiative constant. Because convection in inefficient, the Eddington ratio, $\Gamma_r\simeq\kappa(\rho,T)L_*/(4\pi c G M_*)$, becomes a function of density and temperature only. The Eddington limit (i.e., $\Gamma_r=1$) is a contour of constant opacity in $(\rho,T)$-space,
\beq
\kappa(\rho,T)\simeq \kappa_{\rm Edd} \equiv\frac{4\pi c G M_*}{L_*};
\label{eq:EddingtonContour}
\eeq
I refer to this as the `Eddington contour'. The Eddington contour shifts to lower densities and gas pressures for higher luminosity-mass ratios, which increases with mass (see \citealt{1992A&A...263..129S}, \citealt{1989A&A...210...93L}, or Appendix\,A). The dashed lines in Fig.\,\ref{fig:rho_T_ode} shows two solutions of the Eddington contour for $M_*=10$ and 25\,$\msun$ helium stars with solar metallicity. Luminosities are defined using Eq.\,(\ref{eq:empiricalmesa}).

Consider the ratio of Eq.\,(\ref{eq:diff}) and (\ref{eq:hydrostatic}) in the following form:
\beq
\frac{d{\rm log}(\rho)}{d{\rm log}(T)} = \frac{1-\Gamma_r}{\phi\Gamma_r}-1,
\label{eq:ode_static}
\eeq
where $\phi\equiv p_g/4p_r = \beta/[4(1-\beta)]$. Since the right hand side is a function of only density and temperature, the envelope structure around the iron bump is well-approximated by an ordinary differential equation in one variable, $\rho(T)$. Fig.\,\ref{fig:rho_T_ode} shows solutions to Eq.\,(\ref{eq:ode_static}) for a range of densities starting at $T=10^5$\,K. These solutions are qualitatively similar to Fig.\,10 of G12 which shows envelope solutions for various outer boundary conditions. To understand the variety of solutions, let us consider Eq.\,(\ref{eq:ode_static}) in more detail.

Suppose the envelope solution does not approach the Eddington contour for $T\gtrsim10^{5.2}$\,K and remains sub-Eddington. Near the iron peak, the Eddington contour extends to low densities and, consequently, small $\phi\ll1$. A strongly radiation-dominated, sub-Eddington fluid demands the left side of Eq.\,(\ref{eq:ode_static}) to be very large. As a result, $\rho\rightarrow0$ and an atmosphere forms. These solutions are not shown in Fig.\,\ref{fig:rho_T_ode} since the divergence is extremely rapid and numerically challenging to resolve accurately. 

If the envelope solution instead approaches the Eddington contour then a density inversion forms wherever $\Gamma_r=\Gamma_{\rm inv}\equiv(1+\phi)\inv<1$ \citep{joss}. The opacity increases and the envelope becomes super-Eddington since $\Gamma_{\rm inv}\simeq1-\phi\sim1$. Both the density and Eddington ratio continue to rise until either $\phi$ is not small or the opacity gradient becomes negative: 
\beq
\kappa' \equiv \frac{d{\rm log}(\kappa)}{d{\rm log}(r)} = \kappa_\rho \rho' + \kappa_T T',
\eeq
where the prime indicates a logarithmic radial derivative, $'=d\mrm{log}/d\mrm{log}(r)$, and,
\beq
\kappa_\rho\equiv \frac{\partial{\rm log}(\kappa)}{\partial{\rm log}(\rho)}, \ \ \ \ \ \kappa_T\equiv \frac{\partial{\rm log}(\kappa)}{\partial{\rm log}(T)}.
\eeq
This set of solutions is shown in Fig.\,\ref{fig:rho_T_ode} as solid lines. 

A third solution may appear to exist where the Eddington ratio is neither small and a density inversion does not form: $0\ll\Gamma_r<\Gamma_{\rm inv}<1$. This is not possible beyond the iron peak since the Eddington contour swings to increasing density. There, the envelope must form a density inversion to continue tracing the Eddington contour, which forces $\Gamma_r\ge\Gamma_{\rm inv}$. In principle, these arguments apply to the hydrogen and helium opacity bumps under the same physical conditions: radiation-dominated, near-Eddington, and effectively radiative envelope. This is not seen in some models of massive main-sequence stars by \cite{sanyal15} since convection is not a negligible energy transport mechanism.

\subsubsection{An Approximate Envelope Solution}
\label{sec:unstableenvelopesolutions}

Fig.\,\ref{fig:rho_T_ode} shows the structure of a hydrostatic, radiation-dominated, near-Eddington envelope with inefficient convective transport is well-approximated by the Eddington contour (dashed lines). Because the density, $\rho_{\rm Edd}(T)$, is a function of temperature along the Eddington contour (see Eq.\,\ref{eq:EddingtonContour}), the temperature structure can be approximated for with the integration of Eq.\,(\ref{eq:diff}) given the radius, $R_*$, and temperature $T(R_*)$, at an inner boundary:
\beq
r \simeq R_*\left(1- R_*\int_{T(R_*)}^{T(r)}\frac{4 a_r T^3}{3 G M_* \rho_{\rm Edd}(T) }dT \right)\inv.
\label{eq:radiusestimate}
\eeq
Note that $\Gamma_r$ is one along the Eddington contour. Eq.\,(\ref{eq:EddingtonContour}) and (\ref{eq:radiusestimate}) provide a solution for the density structure $\rho_{\rm Edd}(r)$. 

\begin{figure}
    \centering
    \includegraphics[width=\columnwidth]{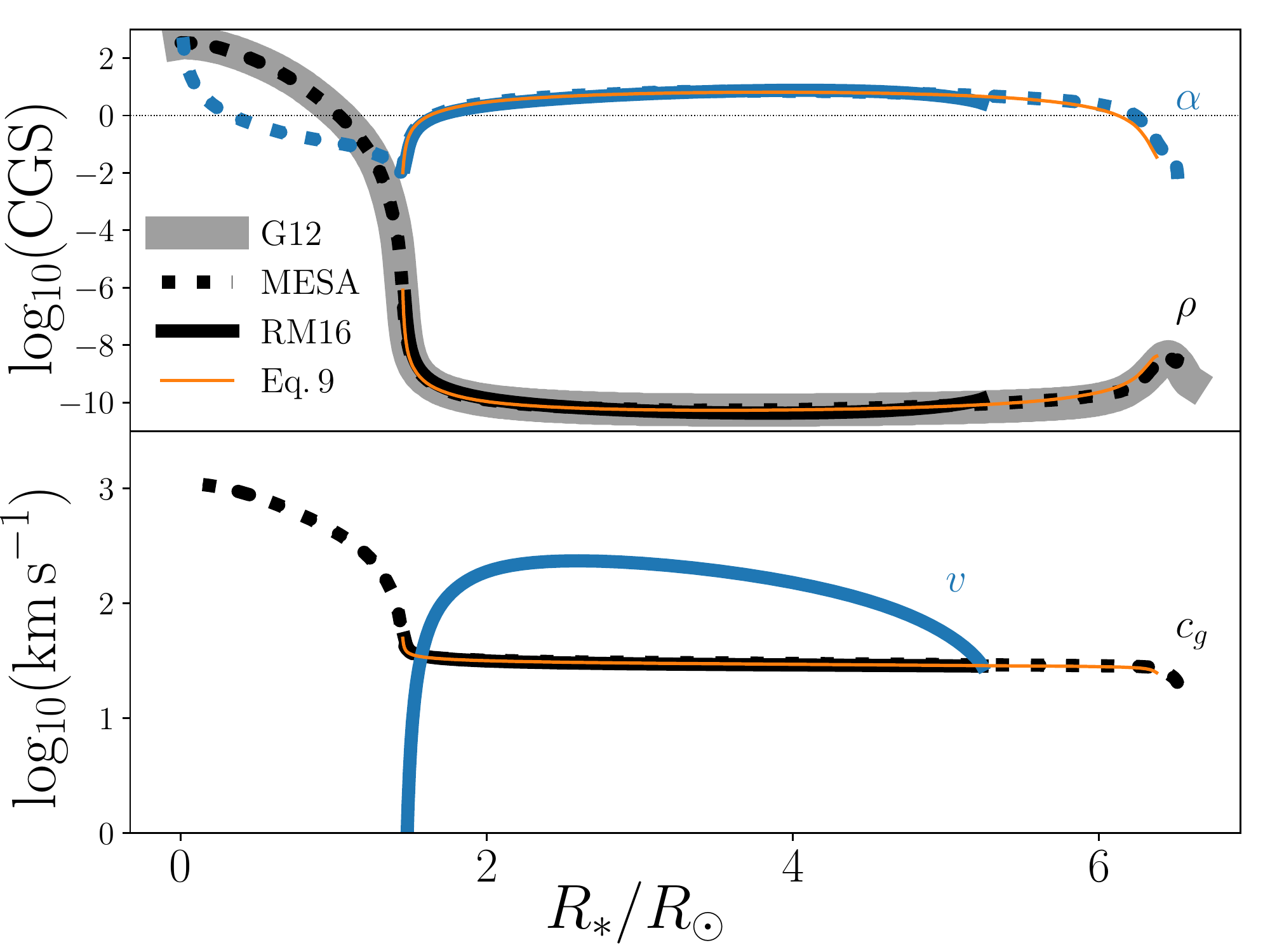}
    \caption{Stellar profiles of a 23 $\msun$ helium stellar model with solar metallicity from G12 (grey) and \texttt{MESA} (dotted). The envelope is extended once the temperature scale height, or radiative energy density, is large: $\alpha\ge1$. Black solid lines are wind models by RM16 for $\mdot=10^{-5.2}\,\msun$\,yr$\inv<\mdot_b$. Thin orange lines are derived from Eq.\,(\ref{eq:EddingtonContour}) and (\ref{eq:radiusestimate}).  }
    \label{fig:bubble}
\end{figure}

Fig.\,\ref{fig:bubble} shows hydrostatic solutions for a 23\,$\msun$ pure helium star with solar metallicity from \texttt{MESA} and G12. A \texttt{MESA} \texttt{inlist} is available in Appendix\,B. The core radius is defined where $R_*=r(T=10^{5.6}\,{\rm K})$ in \texttt{MESA}. Also shown is a steady wind solution with $\mdot=10^{-5.2}$\,\msunpyr from RM16. RM16 categorize this wind solution as `weak' and `extended' since the inertial terms do not affect the envelope structure significantly. It is clear from Fig.\,\ref{fig:bubble} that Eq.\,(\ref{eq:radiusestimate}) is an excellent approximation for all of these solutions.

\subsubsection{Envelope Inflation}
\label{sec:unboundbubble}
RM16 found their wind solutions do not resemble inflated envelopes so long as the temperature scale height (see Eq.\,\ref{eq:diff}),
\beq
\alpha \equiv \frac{H_T}{r}= -\frac{T}{r}\frac{dr}{dT}  = \frac{4p_r}{\rho\Gamma_r v_k^2},
\label{eq:alpha}
\eeq
remains small (i.e., $\alpha<1$). Here, we examine the quality of $\alpha$ as a measure of envelope inflation.

\begin{figure}
    \centering
    \includegraphics[width=\columnwidth]{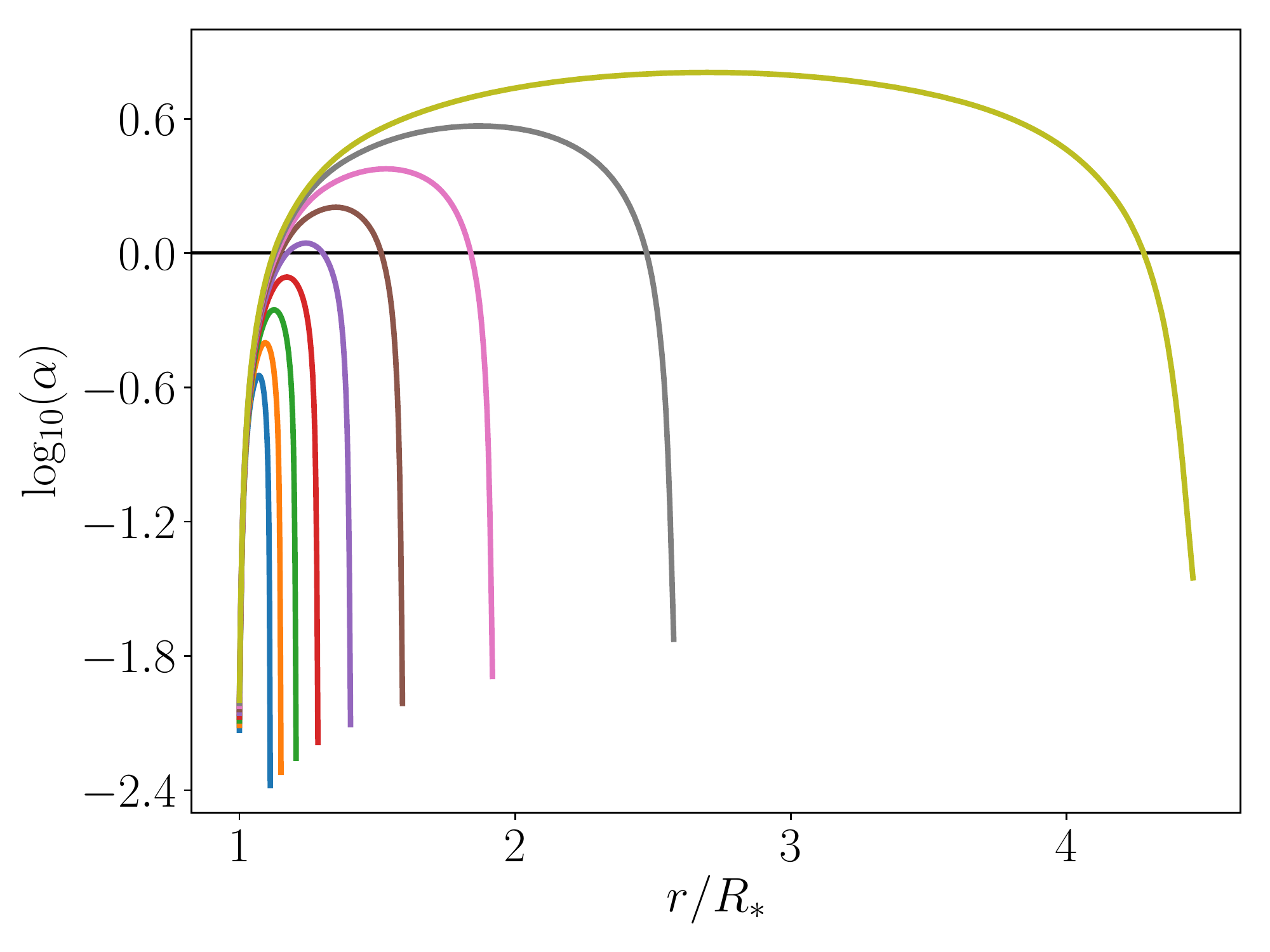}
    \caption{Solutions to Eq.\,(\ref{eq:alpha}) over a temperature domain, $T=10^{5}$\,-\,$10^{5.6}$\,K, for pure helium stellar models with solar metallicity and $M_*=15,16,...,23$\,$\msun$ from \texttt{MESA}. }
    \label{fig:alpha_profile}
\end{figure}

A sequence of $M_*=5-50\,\msun$ hydrogen-free WR stellar models with solar metallicity ($Z_{\odot}=0.02$) and \texttt{GN93hz} abundances are solved for with the stellar evolution code, \texttt{MESA} \citep{2018ApJS..234...34P}. Two chemical compositions are considered to represent WNE and WC interiors: (1) pure helium ($Y=1-Z$, $\mu=4/3$), and (2) a mixture of carbon, oxygen, and helium ($X_{\rm C}+X_{\rm O}+Y=1-Z$, $X_{\rm C}=0.4$, $X_{\rm O}=0.1$, $\mu\simeq1.5$). I construct a second sequence of lower metallicity models ($Z_{\odot}=0.01$) to represent WNE and WC stars in the LMC. Empirical mass-radius-luminosity relations are constructed from these models at temperatures $T=10^{5.6}$\,K well below the iron bump which are available in Appendix\,A. Solutions for Eq.\,(\ref{eq:alpha}) are shown in Fig.\,\ref{fig:alpha_profile} for pure helium models with solar metallicity between $M_*=$15\,-\,23\,$\msun$ and fixed temperature domains, $T=10^{5}$\,-\,$10^{5.6}$\,K. These solutions are found using Eq.\,(\ref{eq:EddingtonContour}) and (\ref{eq:radiusestimate}) with the empirical relations as inner boundary conditions. 

Inferred radii from observed WR stars are found to exceed models without envelope inflation by factors of 3 or more (i.e., $R_{20}/R_*\gtrsim3$). WR models with max($\alpha)>1$ have radii $R_*\gtrsim1.4\,\rsun$ at $T=10^5\,$K. If stellar models with max($\alpha)>1$ are heuristically defined to be `inflated' then the definition is conservatively small in regards to the radius discrepancy problem. Fig.\,\ref{fig:alpha_profile} shows helium models with solar metallicity above $M_b\simeq19\,\msun$ or luminosity $L_b\simeq10^{5.7}\,\lsun$ are considered to be inflated here. Using the relations from \cite{1992A&A...263..129S} instead of those in Appendix\,A gives a minimum mass of $M_b\simeq 14\,\msun$ as found by RM16. The critical mass for envelope inflation is sensitive to the empirical relations despite how similar they may appear (e.g., Fig.\,\ref{fig:mesa}). A criterion for envelope inflation is introduced by \cite{2018A&A...614A..86G}; although, a precise definition is not necessary for the purposes of this investigation.

\begin{figure}[h]
    \centering
    \includegraphics[width=\columnwidth]{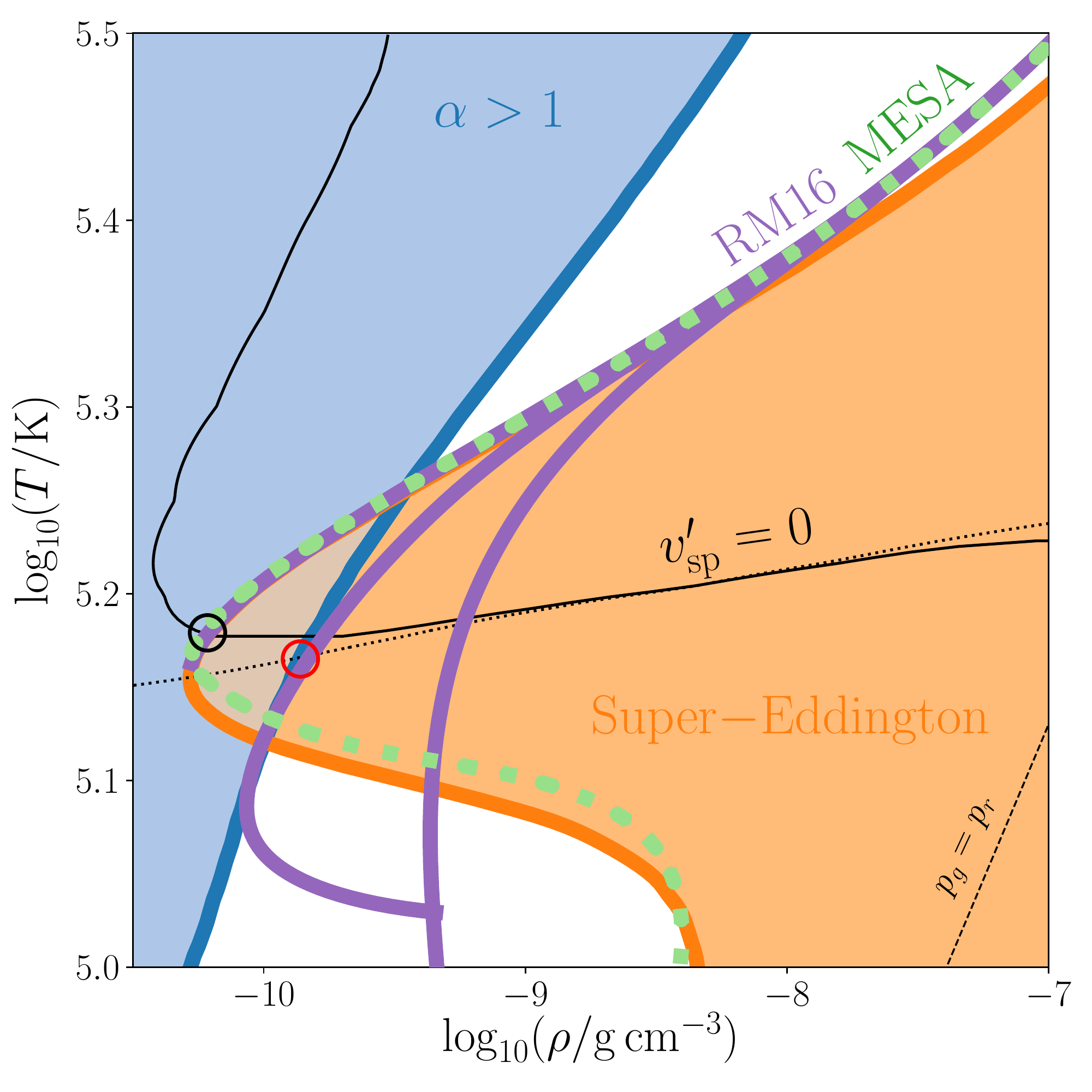}
    \caption{The density-temperature structure around the iron bump for a 23\,$\msun$ stellar model with pure helium and solar metallicity composition. Wind solutions from RM16 are shown with $\mdot$ greater than, less than, and equal to $\mdot_b$ (purple lines). The dashed green line is a hydrostatic \texttt{MESA} solution. The envelope or wind is super-Eddington (inflated) in the orange (blue) domain. Sonic points reside very close to the Eddington contour (orange line) and correspond to wind solutions (i.e., $v'_{\rm sp}>0$; Eq.\,\ref{eq:vprime_b}) when above the solid black line towards larger densities and temperatures. The minimum sonic point density and temperature is located at the intersection (black circle) of the black and orange lines. Eq.\,\ref{eq:vprime_inflation} is evaluated at the red circle at the opacity peak (dotted black line) where $\alpha=1$ (blue line). 
    }
    \label{fig:rhoT}
\end{figure}

\subsection{Dynamic Response} \label{sec:structure_equations}

Consider a steady, spherically symmetric wind,
\beq
\frac{d}{dr}(r^2\rho v) = 0,
\label{eq:steady}
\eeq
with a constant rate of mass loss rate, $\mdot=4\pi r^2 \rho v$, where $v$ is the radial wind velocity. We assume the gas and radiation are in thermal equilibrium and the wind optical depth is sufficiently large to apply the radiative diffusion approximation. The opacity in Eq.\,(\ref{eq:diff}) can be approximated for with the Rosseland mean opacity $\kappa_R(\rho,T)$; the limitations of this are discussed in the next section (\S\ref{sec:validity}) and by RM16. 

Combining Eq.\,(\ref{eq:diff}) and (\ref{eq:steady}) with the momentum equation,
\beq
v\frac{dv}{dr} + \frac{1}{\rho}\frac{d p_g}{dr} =  - \frac{v_k^2(1-\Gamma_r)}{r},
\label{eq:mom1}
\eeq
leads to a wind equation,
\beq
v' = \frac{v_k^2\left(q_g +  \Gamma_r(1+\phi) - 1 \right)}{v^2-c_g^2},
\label{eq:mom2}
\eeq
with dimensionless variable $q_g\equiv2c_g^2/v_k^2\ll1$, gas sound speed $c_g^2=p_g/\rho = k_b T/\mu$, Boltzmann constant $k_b$, and mean molecular weight $\mu$. 

\subsubsection{Sonic Point Conditions}
\label{sec:sonicpoint}
The wind equation is critical where the flow speed matches the gas sound speed, 
\beq 
v(\Rsp)=c_g(\Rsp)\simeq33\,{\rm km\,s}\inv, 
\label{eq:sp1}
\eeq
rather than the total sound speed, $c_s^2=\gamma p/\rho\simeq4p_r/(3\rho)$, where $\gamma$ is the adiabatic index. The location of the sonic point, $\Rsp$, is found where the numerator of Eq.\,(\ref{eq:mom2}) is zero:
\beq
\Gamma_r(\Rsp) = \frac{1 - q_g}{1 + \phi}<1. 
\label{eq:sp2}
\eeq

For time-steady flow, energy conservation states
\beq
\dot{E} = L_r + \mdot \mathcal{B},
\label{eq:edot}
\eeq
where $\dot{E}$ is the rate of stellar energy loss, $\mathcal{B}=w+v^2/2-v_k^2$ is the Bernoulli factor, and
\beq
w = \frac{5}{2}\frac{p_g}{\rho} + \frac{4p_r}{\rho},
\label{eq:enthalpy}
\eeq
is the specific enthalpy. The radiative luminosity is approximately constant near WR sonic points since gas pressure is negligible and $\mdot\mathcal{B}\ll L_*$ (RM16). Therefore, a constant radiative luminosity, $L_r\simeq L_*$, is a good approximation. This simplifies Eq.\,(\ref{eq:sp2}) to become
\beq
\kappa_{\rm sp} \equiv \frac{4\pi c G M_*(1-q_{g,{\rm sp}})}{L_*(1+\phi_{\rm sp})} \simeq \kappa_{\rm Edd},
\label{eq:SPContour}
\eeq
which is a contour in ($\rho,T$)-space in close proximity to the Eddington contour (Eq.\,\ref{eq:EddingtonContour}) since $q_{g,{\rm sp}},\phi\ll1$.

\subsubsection{Minimum Mass Loss Relation for a Transonic Wind: $\mdot_{\rm sp}$}
\label{sec:minimummdotsp}
A wind solution requires the velocity gradient to be positive at the sonic point ($v'_{\rm sp}>0$). RM16 use l'H$\ohat$pital's rule to solve Eq.\,(\ref{eq:mom2}) exactly:
\bea
6\Psi^2 &-& 4\Psi + 1 - 2{\cal W}\left({k_\rho}+\Psi{k_T}\right) \nonumber \\ 
&-&\xi {\cal W}(3q_g\Psi + 8q_g+8q_r -6) > 0 ,
\label{eq:vprime_b}
\eea
where,
\begin{equation*}
\xi = \frac{\mdot v_k^2}{2L_r},~~
{\cal W} = \frac{1-q_g}{q_g},~~
\Psi = \frac{\phi }{1+\phi}{\cal W},
\end{equation*}
\begin{equation*}
\kappa_{\rho} = \frac{d\mrm{log}(\kappa)}{d\mrm{log}(\rho)},~~~~
\kappa_T = \frac{d\mrm{log}(\kappa)}{d\mrm{log}(T)}
\end{equation*}
and,
\begin{equation*}
q_g = \frac{2p_g}{\rho v_k^2},~~
q_r = \frac{2p_r}{\rho v_k^2}\simeq\frac{\alpha}{2},~~
\phi = \frac{p_g}{4p_r},
\end{equation*}
are evaluated at the sonic point. WR winds are far from photon tiring limit at the sonic point and, so, $\xi\simeq0$. Since $\phi$, $q_g$, ${\cal W}\inv$, and $\Psi\ll1$ are small in the domain of interest, Eq.\,(\ref{eq:vprime_b}) is well approximated by the expression 
\beq
\kappa_{\rho}+\Psi\kappa_T<0. 
\label{eq:vprime_b2}
\eeq
The roots of Eq.\,(\ref{eq:vprime_b2}) represent a contour in $\rho$ and $T$ for a given opacity table. This contour is shown as the black line in Fig.\,\ref{fig:rhoT} along with a black dotted contour of the iron peak (i.e., $\kappa_T=0$). Also shown is the Eddington contour for a 23\,$\msun$ with $L_*=10^{5.795}$\,$\lsun$ from RM16. As argued by \cite{nl02}, the sonic point must reside on the hot side of the iron peak to drive a wind. This is not true for sufficiently small sonic point densities due to the factor of $\Psi\propto\alpha\inv\propto\rho$. Instead, Eq.\,(\ref{eq:vprime_b2}) reaches a minimum density of log$_{10}(\rho\,/{\rm g\,cm^{-3}})\simeq-10.5$ and follows a path of nearly constant ratio of radiation-to-gas pressure or $\alpha\sim3$. This feature is relevant for WR stellar models with inflated envelopes where $\alpha>1$.

Given the stellar mass, luminosity and radius relations and an opacity table (i.e., chemical composition), the intersection of contours defined by Eq.\,(\ref{eq:SPContour}) and (\ref{eq:vprime_b}) provide the minimum sonic point density, $\rho_{\rm sp}$, and temperature (or sound speed, $v_{\rm sp}$). The intersection is shown as a black circle in Fig.\,\ref{fig:rhoT}. \cite{1999isw..book.....L} argues the hydrostatic solution (i.e., $\mdot=0$) is a good approximation for the subsonic wind structure. Numerical solutions by \citeauthor{petrovic} and RM16 show this to be accurate in WR winds. Therefore, Eq.\,(\ref{eq:radiusestimate}) can be used to provide an estimate the sonic point radius, $\Rsp$, at the minimum sonic point density or temperature. Combining $\rho_{\rm sp}$, $v_{\rm sp}$, and $\Rsp$ provides an estimate for the minimum mass loss rate for a WR wind to harbour a sonic point:
\beq
\mdot_{\rm sp} \equiv 4 \pi \Rsp^2 \rho_{\rm sp}v_{\rm sp}.
\label{eq:mdotmin}
\eeq
For comparison, \cite{phdthesis} and \cite{2018A&A...614A..86G} assume $\kappa_T=0$ at the minimum sonic point for wind-like solutions (i.e., $v'_{\rm sp}\ge0$). \cite{2018A&A...614A..86G} compute hydrodynamic stellar structure models to support their approximation of the sonic point radius as $R_{\rm sp}\simeq1$\,$\rsun$. \cite{phdthesis} use the empirical mass-radius relation from \cite{1992A&A...263..129S} for the sonic point radius (i.e., $R_{\rm sp}\simeq R_*)$. 

I find the approximations described above are fine for WNE stellar models with masses below $M_*\lesssim19\,\msun$ (or $L_*\lesssim 10^{5.7}\,\lsun$) and solar metallicity. Above these masses (or luminosities), however, the effects from envelope inflation and additional term, $\kappa_\rho$, in Eq.\,(\ref{eq:vprime_b2}) are important. For instance, the previous minimum sonic point condition, $\kappa_T=0$, no longer corresponds to wind-like solutions. Instead, the solutions decelerate at the sonic point and are subsonic at larger radii. These effects also cause Eq.\,(\ref{eq:mdotmin}) to invert with stellar mass which suggests there is a global minimum mass loss relation for WNE stellar models harbouring sonic points at the iron bump; a discussion on this is in \S\ref{sec:globalminmdot}.

\subsubsection{ Maximum Mass Loss Rate for Extended Winds}
\label{sec:bubblemdot}


Fig.\,\ref{fig:rhoT} shows a \texttt{MESA} model of an (hydrostatic) inflated envelope along with three wind solutions from RM16: a compact ($\mdot>\mdot_b$), extended ($\mdot<\mdot_b$), and critical wind ($\mdot=\mdot_b$), where
\beq
\mdot_b \equiv 4\pi R_b^2 \rho_b v_b.
\label{eq:mdotbubble}
\eeq
RM16 found their extended wind solutions resembled inflated envelopes so long as max($\alpha)\geq1$. They solve the subsonic wind is structure between an inner boundary at $T=10^{5.6}\,$K, where the stellar properties are defined by \cite{1992A&A...263..129S}, and the outer boundary at the sonic point. Integration from the sonic point outwards gives the supersonic wind solution, which truncates when the flow is either subsonic again or reaches the He-bump.

To estimate the critical mass loss rate, $\mdot_b$, separating compact and extended wind solutions, RM16 use the density, $\rho_b$, and temperature, $T_b$, where $\alpha=1$ at the iron peak (red circle in Fig.\,\ref{fig:rhoT}). The critical wind solution is approximately tangential to the $\alpha=1$ contour (i.e., $\rho\simeq 4T'$) which implies the velocity gradient follows
\beq
v'_b =  - T'\left(\frac{\rho'}{T'} \right) - 2 \simeq \frac{4}{\alpha\Gamma_r}-2, 
\label{eq:vprime_inflation}
\eeq
using Eq.\,(\ref{eq:alpha}) and (\ref{eq:steady}). With Eq.\,(\ref{eq:mom2}), the supersonic velocity at the iron peak is then
\beq
v_b^2 \simeq c_g^2 + \frac{2c_g^2 + v_k^2\left(\Gamma_r(1+\phi)-1\right)}{4(\alpha\Gamma_r)\inv -2}.
\label{eq:supersonicv}
\eeq
Since the temperature scale height is small where max($\alpha)\leq1$, the stellar radius, $R_*=r(T=10^{5.6{\rm\,K}})$ or Eq.\,(\ref{eq:empiricalmesa}) is taken for $R_b$. RM16 found the approximations for the critical mass loss relation $\mdot_b$ accurately distinguish their WR wind solutions.

The estimate for $\mdot_b$ is based on the supersonic conditions of the outflow where $c_g\lesssim v_b \lesssim 300$\,km\,s$\inv\sim(0.1-0.2)v_{\infty}$. The Rosseland mean approximation certainly fails at this point and Doppler effects become important. The neglect of radiative forces besides continuum acceleration implies $\mdot_b$ is an overestimate. In other words, envelope inflation, as defined in \S\ref{sec:unboundbubble}, is likely erased for mass loss rates smaller than $\mdot_b$ once all acceleration mechanisms are accounted for. 

\subsection{Validity of Approximations}\label{sec:validity}

\subsubsection{Radiative Envelope Structure}
\label{sec:convection}
The dynamic condition for convective instability is (RM16)
\beq
\Gamma_r \ge  \left(1 + \frac{v^2}{v_k^2}v' \right) \Gamma_c. 
\label{eq:convection2}
\eeq
RM16 found little acceleration is necessary to suppress convection in a radiation-dominated envelope since $\Gamma_c\simeq1$ (Eq.\,\ref{eq:convection}). Because this study considers solutions where $v'_{\rm sp}=0$, the convective instability criterion is satisfied in some region in the subsonic domain for the slowest winds. The minimum mass loss rate to suppress subsonic convection (solving Eq.\,\ref{eq:mdotmin} with \ref{eq:convection}, \ref{eq:mom2}, and \ref{eq:convection2}) 
differs from $\mdot_{\rm sp}$ by a few percent. Therefore, convection is not important for studies of WR sonic points. 

\cite{yfjiang15} find massive hydrogen envelopes with a subsurface iron bump are inherently unstable to radiatively-driven hydrodynamic instabilities (RHIs; \citealt{2003ApJ...596..509B}). Their \texttt{StarTop} model shows the density inversion from a hydrostatic one-dimensional model initially fragments, collapses, and washes out in multi-dimensional hydrodynamic simulations with radiative transfer. The authors found non-adiabatic mixing length models significantly overestimate the convective flux and find the effects of porosity and vertical oscillations to be important. While turbulent motions can transport kinetic energy, the flux remains small since the motions are comparable to the gas sound speed. The radiative acceleration remains larger than the gravitational acceleration and, as a result, a time- and spatially-averaged density inversion manifests at later times.

RM16 argue the criterion for RHIs is satisfied beneath the iron peak in hydrostatic WR envelopes around $T\sim10^{5.6}$\,K. In stellar oscillation theory, these motions are referred to as `strange' modes \citep{1990MNRAS.245..597G,2003ApJ...596..509B} which have been found to excite pulsations in quasi-static helium envelopes models \citep{1993MNRAS.262L...7G, 1994MNRAS.271...66G, gras16}. Turbulent energy transport is inefficient once the optical depth per pressure scale height, $\tau_0=\kappa\rho H$, is less than $\tau_b=c/c_g$ \citep{yfjiang15}. Since $H\simeq H_T/4$ in a radiation-dominated envelope, we have 
\bea
\frac{\tau_0}{\tau_b} &\simeq& \kappa \rho \frac{H_T}{4}\left(\frac{c}{c_g}\right)\inv \nonumber\\
&\simeq& \frac{\alpha GM_*\mdot}{L_*R_*} \nonumber \\
&\simeq& 0.07 \alpha\left(\frac{M_*}{10\msun} \right)^{-1.3}\left(\frac{\mdot}{10^{-4}\msun{\rm \, yr}\inv}\right),
\eea
using Eq.\,(\ref{eq:EddingtonContour}), (\ref{eq:alpha}), (\ref{eq:steady}), (\ref{eq:empiricalmesa}) and (\ref{eq:empiricalmesa2}). Because $\tau_0/\tau_b$ is very small for typical WR parameters, the envelopes at the iron bump resemble the \texttt{StarTop} model by \cite{yfjiang15}. Therefore, turbulent energy transport is not expected to be important in WR envelopes and the adiabatic estimate for convective transport, Eq.\,(\ref{eq:max_lconv}), is likely overestimated. This further supports the argument that convection is not energetically relevant around the iron bump in WR stars.

\subsubsection{Rosseland Opacity and Microturbulence}
\label{sec:microturbulence}

\cite{nl02} and RM16 argue the Doppler shift from flow acceleration is small in comparison to the iron line widths across the sonic point. 
They apply the Rosseland mean approximation in their wind calculations so long as the optical depth parameter \citep{1975ApJ...195..157C} is large,
\beq
t_{\rm CAK} = \frac{\sigma_e v_{\rm th} \rho}{dv/dr}\gtrsim1,
\label{eq:tcak}
\eeq
where $\sigma_e=0.2\,$cm$^2$\,g$\inv$ is the electron scattering opacity of helium. A problem with the authors' argument is the referenced thermal width, $v_{\rm th}=\sqrt{2}c_g$, is for hydrogen and not iron. Using the iron thermal width, $v_{\rm th,Fe}\simeq7$\,km\,s$\inv$, reduces $t_{\rm CAK}\propto \mu^{-1/2}$ by about $7.5$. For weak or low density winds $\mdot\lesssim10^{-5}\,\msun\,$yr$\inv$, the Rosseland approximation degrades close to the sonic point but not significantly: $\mathrm{min}(t_{\rm CAK})\gtrsim0.3$, in comparison to O/B stellar atmospheres/winds where $t_{\rm CAK,O}\simeq10^{-2}$.

Turbulent motions from sub-surface convection \citep{gras16} can broaden photospheric iron lines \citep{2009A&A...499..279C}, increase $\mathrm{min}(t_{\rm CAK})$ and reduce the effects from Doppler motion. As discussed in \S\ref{sec:convection}, WR envelopes are susceptible to RHIs and may not be time-steady on scales smaller than the local scale height. If these motions are acoustic (barring non-linear acoustic effects), the photon mean free path, $\ell_{\rm ph}\sim(\kappa_{\rm Edd} \rho)\inv\propto \rho\inv$, increases more rapidly than the wavelength, $\lambda \propto c_s\propto\rho^{1/6}$ ($\lambda\propto\rho^{-1/2}$ for isothermal gas), from high to low temperatures. Therefore, inhomogeneities from RHIs are bound to become optically thin and become highly compressible. Indeed, \cite{yfjiang15} see strong shock formation in their \texttt{StarTop} models.

\cite{yfjiang15} show the inhomogeneous turbulent gas is radiatively porous in their \texttt{StarTop} models as photons preferentially flow through channels of smaller relative optical depth. The reduction in the effective radiative acceleration from porosity affects the stellar structure, although, weakly. These calculations employ the OPAL opacity tables which assumes the fluid is motionless on scales smaller than the photon mean free path. If WR envelopes harbour small-scale or `microturbulent' motions then the desaturation of optically thick lines may increase the amount of line overlap, the total opacity, and effective radiative acceleration. Microturbulence may be an important factor in determining the structures of outer WR envelopes and winds. To employ the Rosseland mean approximation in the subsonic domain of weak winds, microturbulence must be the dominant contributor to the total line width of iron, $v_{\rm th,Fe}/[\mathrm{min}(t_{\rm CAK})]\sim23$\,km\,s$\inv\simeq0.7c_g$, such that $t_{\rm CAK}\gtrsim1$. This is not a restrictive requirement considering the hydrostatic iron line widths are small and subsonic $v_{\rm th,Fe}\simeq 7$\,km\,s$\inv\simeq0.2 c_g$. 

In \S\ref{sec:results}, I show that an opacity enhancement is necessary for WR winds to have sonic points at the iron bump. Because opacity tables do not include the effects from microturbulent broadening, I use opacity tables with enhanced metallicity as a rough approximation for this effect. While a metallicity enhancement includes opacity contributions from other elements, iron is by far the dominant opacity source at this temperature regime. Implicit in these approximations is the assumption that the turbulent gas can be described using time-steady calculations. This is not justifiable since there are currently no multidimensional simulations of WR outflows.





\subsubsection{Enhanced Iron Opacity Tables}
\label{sec:laboratory}
Rosseland mean opacity tables are found to underestimate the contributions from iron under solar tachocline conditions ($T\simeq2\times10^{6}$\,K,\,$n_e\simeq10^{22-22.5}$\,cm$^{-3}$) in direct laboratory measurements by 55-75\% \citep{2015Natur.517...56B}. Half of the helioseismic discrepancy is resolved if this correction is included. This discrepancy may be eliminated once other elements (e.g., Ni) are included. Corrected regions of stellar instability no longer show a discrepancy for the log\,$T_{\rm eff}$\,vs.\,log\,$g$ locations of hybrid slowly pulsating B(SPB)-$\beta$ Cephei pulsators \citep{2016MNRAS.455L..67M}. 

The conditions for wind and inflated envelope formation in WR stellar models are strongly sensitive to the iron opacity. While the iron bump considered here is at cooler temperatures than laboratory conditions, it has several more bound-bound and bound-free transitions available and, so, opacity enhancement corrections are possible.

\section{Results}\label{sec:results}
In summary, there are two critical mass loss relations for a WR stellar model to harbour a sonic point, $\mdot_{\rm sp}(L_*)$, and inflated envelope, $\mdot_b(L_*)$. These relations partition ($\mdot,L_*$)-space into four regions that correspond to the following responses to the iron bump:
\begin{enumerate}
\itemsep0em 
    \item Compact envelope (no wind): $\mdot<\mdot_{\rm sp}$ and $L_*<L_b$
    \item Inflated envelope (no wind): $\mdot<\mdot_{\rm sp}$,   $L_*>L_b$
    \item Compact wind:  $\mdot>\mdot_{\rm sp}$ and $\mdot>\mdot_b(L_*)$
    \item Extended wind:  $\mdot>\mdot_{\rm sp}$ and $\mdot<\mdot_b(L_*)$
\end{enumerate}
Note that WR models with $L_*\lesssim L_b$ have max$(\alpha)<1$ and do not have inflated envelopes as defined in \S\ref{sec:unboundbubble}. Therefore, the mass loss rate necessary to erase an inflated envelope is definitively zero: $\mdot_b(L_*<L_b)=0$.

\begin{figure*}
    \centering
    \includegraphics[width=\textwidth]{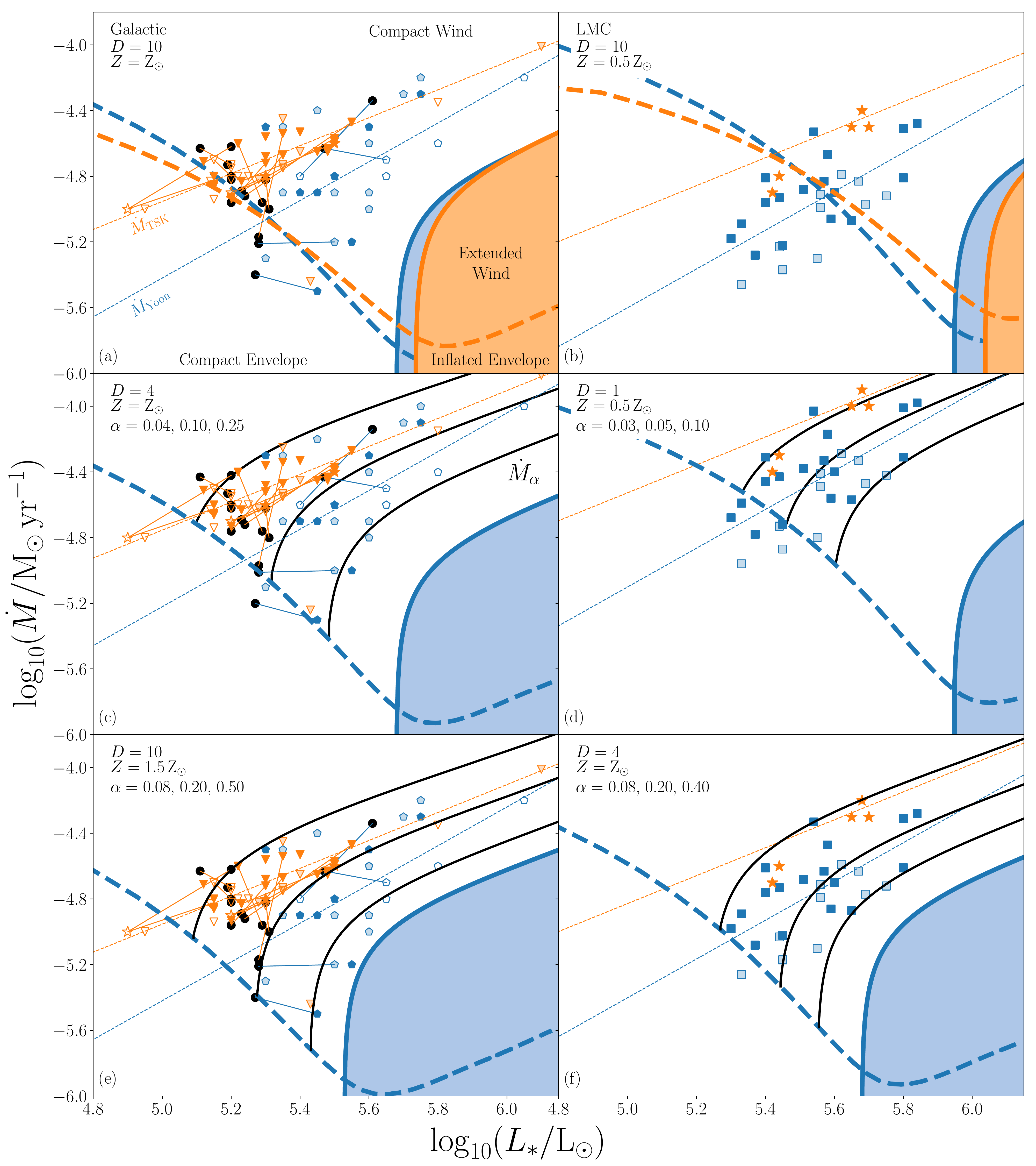}
    \caption{Mass loss distribution of of WNE (blue) and WC (orange) hydrogen-free single stars from the MW (left panels) and LMC (right panels) with clumping factor $D$. Redundant WR stars are connected by lines. Filled ($R_{20}/R_*(L_*)<3$), translucent ($3\le R_{20}/R_*(L_*)<5$), or unfilled points ($R_{20}/R_*(L_*)\ge5$) correspond to a ratio of observed $R_{20}$ and expected radii $R_*(L_*)$ given the observed luminosity, $L_*$, and empirical relations in Appendix A. Thin dashed lines are empirical mass loss relations from \citep{2016ApJ...833..133T} and \citep{2017MNRAS.470.3970Y}. $\mdot_{\rm sp}$ and $\mdot_b$ are thick dashed and solid lines, respectively, and derived from opacity tables with metallicites, $Z$, listed in the top-left corner. $\mdot_{\alpha}$ are the black solid lines (see \S\ref{sec:theoreticalmdot}) with $\max(\alpha)$ listed in the top-left corner. Points: $\CIRCLE$ ($R_{20}$ not available): \cite{2000A&A...360..227N}, $\pentagon$: \cite{2006A&A...457.1015H}, $\bigtriangledown$: \cite{2012A&A...540A.144S}, $\square$:  \cite{2014A&A...565A..27H}, and $\bigwhitestar$:  \cite{2015A&A...581A.110T}.}
    \label{fig:mdot_lum}
\end{figure*}

Figures\,\ref{fig:mdot_lum}a,b show the critical mass loss relations for metallicity abundances in the Milky Way (MW) and Large Magellanic Cloud (LMC), respectively, and WNE and WC chemical compositions (\S\ref{sec:unboundbubble}). Also shown are observations of single hydrogen-free WNE and WC stars \citep{2000A&A...360..227N,2006A&A...457.1015H,2012A&A...540A.144S,2014A&A...565A..27H, 2015A&A...581A.110T, 2016ApJ...833..133T} along with empirical mass loss relations from \cite{2016ApJ...833..133T}, $\mdot_{\rm TSK}$, and \cite{2017MNRAS.470.3970Y}, $\mdot_{\rm Yoon}$. The rates from \cite{2006A&A...457.1015H} are adjusted so the clumping factor is $D=10$, instead of 4, to homogenize the data set \citep{2017MNRAS.470.3970Y}. Redundant observations between multiple surveys are connected by lines, which provide an impression of the systematic errors. Uncertainties are cited to be of order $\Delta[\mdot,L_{\rm LMC},L_{\rm MW}]\sim[0.2, 0.2, 0.4]$\,dex. 
Non-solid markers are WR stars with inferred radii, $R_{20}$, above a factor of three larger than predicted from hydrostatic models without envelope inflation at the same total luminosity. The chemical composition is found to have a relatively small influence. I use the critical mass loss elation from WNE stellar models in the remaining panels in Fig.\,\ref{fig:mdot_lum}. 

\section{Discussion}
\label{sec:discussion}
\subsection{Are Wolf-Rayet Envelopes Inflated?}
\label{sec:arewrstarsinflated}
Models of MW and LMC WR stars are found to have either sufficiently small luminosities, $L_*<L_b$, or large mass loss rates, $\mdot>\mdot_b$, to prevent the formation of an extended wind (i.e., envelope inflation) at the iron bump. Therefore, the hydrostatic approximation is not valid for WR models with both envelope inflation and observed rates of mass loss. In other words, hydrostatic WR models with envelope inflation and empirical rates of mass loss are not consistent with one-dimensional hydrodynamic stellar equations. This is a conservative statement considering the neglect of additional radiative forces (e.g., Doppler shifts, line-deshadowing) and definition of `inflated envelopes' used in this study (see \S\ref{sec:unboundbubble}). Decreasing the clumping factor from $D=10$ increases the observed mass loss rates and further degrades the hydrostatic approximation.

LMC WRs are observed to have slightly smaller mass loss rates and large inferred radii. These observations could be described by an extended wind with a larger iron opacity bump. This may be achieved with a substantial increase in either the iron abundance or opacity enhancement from microturbulent motions.

G12 propose macroscopic imhomogeneities as an opacity enhancement mechanism to increase the effects of envelope inflation. An opacity enhancement could also drive stronger outflows which would inhibit envelope inflation. Inhomogeneities in multidimensional simulations from \cite{yfjiang15} are found to render a medium porous to radiation which would deflate an inflated WR envelope. The effects of inhomgeneties remain speculative for now and until multidimensional simulations of WR envelopes arise. Without appeals to substantial opacity enhancement, one-dimensional WR models with observed rates of mass loss do not harbour inflated envelopes and, consequently, the radius discrepancy problem remains outstanding.

\subsection{Global Minimum Mass Loss Rate}
\label{sec:globalminmdot}
Fig.\,\ref{fig:mdot_lum}a,b show $\mdot_{\rm sp}(L_*)$ for WNE and WC stellar models with solar and sub-solar metallicities. Less massive models have smaller $L_*/M_*$ ratios and require larger opacities, densities, and mass loss rates to harbour a sonic point. This is reflected in the rise of $\mdot_{\rm sp}(L_*)$ for smaller $L_*\lesssim10^{5.7}$\,$\lsun$. The $\mdot_{\rm sp}(L_*)$ relation is in agreement with the semi-analytical results by \cite{phdthesis} and \cite{2018A&A...614A..86G} for $L_*\lesssim L_b\sim10^{5.7}$\,$\lsun$ or $M_*\lesssim M_b\sim 19$\,$\msun$. 

Stellar models with $L_*\gtrsim L_b$ develop inflated envelopes and have sonic points at radii larger than the core (i.e., $R_{\rm sp}/R_*\gtrsim1.4$). Furthermore, the minimum sonic point density for a wind-like solution (i.e., $v'_{\rm sp}\ge0$) increases, rather than decreases, with stellar luminosity. As a result, $\mdot_{\rm sp}(L_*)$ increases for $L_*\gtrsim L_b$ and a global minimum mass loss rate emerges at roughly min($\mdot)\simeq10^{-6}\,\msun$\,yr$\inv$. This increases slightly to min($\mdot)\simeq10^{-5.7}\,\msun$\,yr$\inv$ for WR models with LMC metallicities which have $L_b\sim10^6$\,$\lsun$. The discovery of a global minimum mass loss rate for an iron bump to harbour a sonic point is only found when including the effects from envelope inflation and the conditions for wind-like sonic points.

\subsection{Sonic Point Discrepancy}
\label{sec:sonicpointdiscrepancy}
A small fraction of MW WRs in Fig.\,\ref{fig:mdot_lum}a appear to have insufficient mass loss rates or luminosity to be driven by the iron bump (i.e., $\mdot\lesssim\mdot_{\rm sp}(L_*)$). This `discrepancy' is small in comparison to the relatively large observational uncertainties. Therefore, stellar models of observed MW WR stars can harbour sonic points at the iron bump.

If there were a sonic point discrepancy at the iron bump for MW WR stars then a reduced clumping factor from $D_{\rm MW}=10$ to 4 (i.e., increase $\mdot_{\rm MW}$ by $\sqrt{10/4}\simeq1.6$) is one possible resolution (Fig.\,\ref{fig:mdot_lum}c). \cite{2017MNRAS.470.3970Y} suggests such a correction from their evolutionary calculations. The discrepancy can also be explained using an opacity table with 50\% increase in metallicity (Fig.\,\ref{fig:mdot_lum}e). In \S\ref{sec:microturbulence} and \S\ref{sec:laboratory}, I describe two motivations for opacity enhancement regarding microturbulence and laboratory measurements. Employing clumping factors larger than $D_{\rm MW}=10$ or opacity tables with smaller metallicity/opacities for MW WR stars would widen the discrepancy.

Half of LMC WR stars have $\mdot\lesssim\mdot_{\rm sp}$ (Fig.\,\ref{fig:mdot_lum}b). A homogeneous wind (i.e., $D_{\rm LMC}=1$) does not resolve this discrepancy and would be inconsistent with clumped wind observations \citep{2007ARA&A..45..177C,nsmith14}. An empirical clumping factor of $D_{\rm LMC}=4$ requires an opacity enhancement of twice the LMC metallicity $Z\gtrsim 2Z_{\rm LMC}$ (Fig.\,\ref{fig:mdot_lum}f). 

WR models with $\mdot=\mdot_{\rm sp}$ have peculiar `wind' structures. First, their winds experience zero acceleration at the iron peak (i.e., $v'_{\rm sp}=0$) and a maximum velocity that is the gas sound speed, which is very small, $c_g\ll v_{\infty}$. In one-dimension, this suggests the helium opacity bump may drive the wind instead. This could further explain the sonic point discrepancy for LMC WR stars since the density, temperatures, and $\mdot_{\rm sp}$ are smaller at the helium bump than the iron bump. There are issues with the prospect of steady winds launched from the helium bump. For example, a density inversion would remain between the iron and helium bumps. One-dimensional, time-steady calculations are likely inappropriate for this problem considering how density inversions are dynamically unstable in multidimensional simulations \citep{yfjiang15}. RM16 found wind models most wind models with $\mdot>\mdot_{\rm sp}({\rm Fe})$ do not reach escape velocities, but rather decelerate and form a density inversion. If such winds do launch from the helium bump then there exists three sonic points in the velocity field; a scenario where multi-dimensional effects are likely important.

\subsection{WR Mass Loss Relation}
\label{sec:theoreticalmdot}
RM16 found their extended and compact wind models are differentiable if max($\alpha)$ is above or below unity, respectively. A critical mass loss relation for extended winds, $\mdot_b(L_*)$, where max($\alpha)=1$ was derived as a result. Suppose, instead, $\alpha$ did not exceed some smaller critical value than one due to other constraints set at, for example, the photosphere, infinity, or at a critical point like those in line-driven winds \citep{1975ApJ...195..157C}. A new mass loss relation, $\mdot_{\alpha}(L_*)$, can be calculated in a similar fashion to $\mdot_b\equiv\mdot_{\alpha=1}$ for a different value of max($\alpha)$ at the iron peak. Fig.\,\ref{fig:mdot_lum}c-f show $\mdot_{\alpha}(L_*)$ for various critical max($\alpha$) and metallicities. Empirical relations for $\mdot_{\alpha}$ are available in Appendix\,B.

The distribution of mass loss rates from WR stars and $\mdot_{\alpha}(L_*)$ are qualitatively similar. Increasing the metallicity by $\Delta Z = 0.01$ and reducing the clumping factor, $D$, such that $\mdot\gtrsim\mdot_{\rm sp}$ leads to a homogeneous range of $0.1\lesssim$\,max($\alpha$)\,$\lesssim0.5$ that spans both MW and LMC WR distributions (Fig.\,\ref{fig:mdot_lum}e,f). Smaller max$(\alpha)$ values shift the $\mdot_{\alpha}(L_*)$ relation towards smaller luminosities (i.e., $L_{\alpha}<L_b$). WC stars typically have larger $\mdot$ than WNE stars and, so, are better fit with $\mdot_{\alpha}(L_*)$ for smaller max$(\alpha)$. 

The $\mdot_{\alpha}(L_*)$ relation increases rapidly with $L_*$ for less luminous stellar models since their Eddington ratios are closer to unity and supersonic velocities, $v^2\simeq v_k^2(\Gamma_r-1)$, are smaller at the iron peak (see Eq.\,\ref{eq:supersonicv}). The growth is more gradual for higher $L_*$ since $\Gamma_r-1\sim1$. These dependencies lead to a steep and shallow component to $\mdot_{\alpha}(L_*)$, respectively. The minimum mass loss relation to harbour a wind-like sonic point, $\mdot_{\rm sp}$, truncates the steep component for less luminous stellar models. As a result, less luminous stars with larger $\mdot$ have shallower overall gradients in $d\dot{M}/dL_*$ when compared to more luminous stellar models with smaller $\mdot$. This is a noted distinction between WC and WNE populations \citep{vink05, 2014A&A...565A..27H, 2016ApJ...833..133T, 2017MNRAS.470.3970Y}. The steep component reduces for larger $\mdot_{\rm sp}$ or smaller metallicities (e.g., Fig.\,\ref{fig:mdot_lum}d and f).

\subsubsection{$\beta$-law Wind Models}
\label{sec:betalaw}
The densities of supersonic winds are certainly small enough to expect other radiative forces not included here. For example, line-deshadowing is responsible for driving optically thin winds from O/B stars which follow a $\beta$-law velocity structure \citep{1975ApJ...195..157C}. \cite{2013A&A...560A...6G} and \cite{2017A&A...608A..34G} use a $\beta$-law as a scaffold between the outer optically thin and inner optically thick wind to calculate the radiative forces. That is, the radiative forces are assumed to drive a $\beta$-law velocity structure. Their outer boundary condition is the ratio of wind speeds at infinity and $R_{20}$. Although, the physical meaning of $R_{20}$ is unclear if it is neither the sonic point at the iron bump, $R_{\rm sp}$, nor the core radius, $R_*$. 
Their wind models generate a mass loss relation that matches the galactic WNE distribution for $D=10$. The disagreement between their relation with the LMC WR distribution may come from the non-existence of a sonic point at the iron bump ($\mdot<\mdot_{\rm sp}$). As the authors suggest, this may further indicate the sonic point resides at the helium opacity bumps. Another resolution is to reduce the clumping factor and an enhance the iron opacity bump.

A prediction from \cite{2013A&A...560A...6G}, \cite{2017A&A...608A..34G}, and \cite{2018ApJ...852..126N} is a nearly constant ratio of radiation and gas pressure $p_r/p_g\sim100-160$ in the supersonic wind. Fig.\,\ref{fig:mdot_lum}c,d suggest galactic WNE winds are best fit with $\mdot_{\alpha}(L_*)$ for max($\alpha)\sim0.2$. This sets a ratio of radiation to gas pressure, 
\beq
\frac{p_r}{p_g} = \frac{v_k^2\times{\rm max}(\alpha) }{4 c_g^2} \simeq 145\left( \frac{{\rm max}(\alpha) }{0.2}\right)\left(\frac{M_*}{15\msun}\right)^{0.42},
\label{eq:criticalradgaspressure}
\eeq 
at the iron peak using our empirical WR relations (Appendix\,A) and Eq.\,(\ref{eq:alpha}). This ratio is in agreement with the mentioned authors' findings despite the different approaches to our wind model calculations. Envelope inflation, as defined in \S\ref{sec:unboundbubble}, is not consistent with their models considering $p_r/p_g\gtrsim700$ for max$(\alpha)\gtrsim1$. Our models suggests max$(\alpha)$ and $p_r/p_g$ in WC winds is about one-half of that in WNE winds.

\section{Summary}\label{sec:conclusion}
This investigation explores several aspects regarding the structure of WR stellar models at the iron bump. First, a radiation-dominated, radiative, near-Eddington and hydrostatic envelope must develop either an atmosphere or density inversion at a prominent opacity bump (\S\ref{sec:bubbles}). The envelope structure is primarily determined by the Eddington contour (\S\ref{sec:unstableenvelopesolutions}), which depends on the $L_*/M_*$ ratio and the chemical composition. Models with larger $L_*/M_*$ or iron abundance have Eddington contours at smaller densities which increases both the temperature scale height, $H_T\propto\rho\inv$, and envelope size. The radius discrepancy problem is the original motivation for this investigation and, so, a heuristic and conservative definition for envelope inflation is adopted: $\alpha\equiv H_T/R_*>1$, where $R_*$ is the radius at $T=10^{5.6}$\,K.

Second, this investigation compares the observed distribution of WR mass loss rates to the critical rate, $\mdot_b$, to harbour extended winds (i.e., inflated envelopes with time-steady, transonic flow) assuming typical wind parameters and opacity tables. I show extended winds do not manifest since the observed WR distribution have either insufficient luminosities or strong winds. This suggests previous hydrostatic WR stellar models with both envelope inflation and mass loss are not consistent with the one-dimensional hydrodynamic stellar structure equations. In other words, the hydrostatic approximation is invalid near the iron bump in WR stellar models. This conclusion is conservative considering the effects from Doppler shifts and microturbulent line broadening, which are expected to increase the radiative acceleration, are not included. 

Envelope inflation is a problematic resolution to the WR radius discrepancy problem for various other reasons. For example, the radius discrepancy problem would not be resolved but, rather, inverted for MW WR stars (i.e., why are some MW WR radii too small?). Another example is that WR wind models suggest the structures have constant radiation-to-gas pressure ratios of $p_r/p_g\simeq100-160$ \citep{2013A&A...560A...6G, 2017A&A...608A..34G,2018ApJ...852..126N}. These values correspond to $\alpha\simeq0.2$ and temperature scale heights (see \S\ref{sec:betalaw}) that are too small for envelope inflation to manifest using the definition in \S\ref{sec:unboundbubble}. WR stellar models with higher metallicities develop inflated envelopes at smaller luminosities. Fig.\,\ref{fig:mdot_lum} suggests more than twice the host environment metallicity is required to resolve the radius discrepancy problem with this explanation.

Third, \cite{phdthesis} and \cite{2018A&A...614A..86G} calculate the minimum mass loss rate for a WR stellar model to harbour a sonic point on the hot side of the iron bump (i.e., $\kappa_T<0$; \citealt{nl02}). Respectively, the authors approximate the sonic point radius with either the mass-radius relation from \cite{1992A&A...263..129S} or a constant value of 1\,$\rsun$. Here, I show these approximations are valid for less massive stellar models without envelope inflation (i.e., $M_*\lesssim19$\,$\msun$). In more massive stellar models, the minimum sonic point density for wind-like solutions is larger than suggested by conditions at the iron peak once the entire sonic point condition is included. The sonic point radius is also larger than previously approximated due to the effects of envelope inflation. This combination is found to increase, rather than decrease, the minimum mass loss relation with luminosity for hydrogen-free WR stars with $M_*\gtrsim19$\,$\msun$ or $L_*\gtrsim10^{5.7}$\,$\lsun$ with solar metallicity. Likewise, $M_*\gtrsim27$\,$\msun$ or $L_*\gtrsim10^{5.9}$\,$\lsun$ for models with LMC metallicities. As a result, hydrogen-free WR stars in the MW and LMC have global minimum mass loss rates of $10^{-6}$ and $10^{-5.7}$\,$\msun$\,yr$\inv$, respectively, if they harbour sonic points at the iron bump.

Fourth, half of early-type WR stars in the LMC have sufficiently small $\mdot$ such that their sonic points do not reside at the iron bump. The discrepancy vanishes given twice the LMC metallicity and a reduced clumping factor from $D=10$ to 4. Without a reduction in the clumping factor, super-solar metallicities are necessary. In this paper, I suggest microturbulence and laboratory corrections may enhance the opacity tables and resolve the sonic point discrepancy. 

Indeed, early-type WR stars with mass loss rates above the minimum, $\mdot_{\rm sp}$, does not guarantee the winds escape via the iron bump. RM16 found nearly all of their wind solutions did not escape due to the steep potential; although, the effects from Doppler shifts were not included. Another scenario is that a subset of WR winds escape via the helium bump instead. That is, WR mass loss is not driven by radiation on UV lines but rather by the continuum field on the helium bump. The viability of this scenario requires an understanding of the deeper structure around the iron bump which may not be describable with classic mixing length theory, existing opacity tables, or one-dimensional time-steady calculations.

Lastly, the critical mass loss relation separating extended and compact winds is found to trace the observed WR distribution when smaller values of ${\rm max}(\alpha)=1$ are considered (\S\ref{sec:theoreticalmdot}). Nominal values of ${\rm max}(\alpha)\simeq0.1-0.2$ correspond to $p_r/p_g\simeq70-145$, which is the typical range of values argued by \cite{2013A&A...560A...6G}, \cite{2017A&A...608A..34G}, and \cite{2018ApJ...852..126N}. If the critical relation is accurate then it suggests the mass loss rate declines faster than a single power-law at late stages of evolution. This may have interesting implications for both the star and its circumstellar environment before core-collapse.

$\\ \\$
SR thanks Chris Matzner, Eliot Quataert, and Yan-Fei Jiang for useful conversations. SR appreciates the referee for their comments and suggestions. This research was funded by the Gordon and Betty Moore Foundation through Grant GBMF5076.

\section*{Appendix\,A\\ Empirical Fits to WR \texttt{MESA} Models}
\label{sec:appendixa}
\begin{figure}[ht]
  \centering
 \includegraphics[width=\columnwidth]{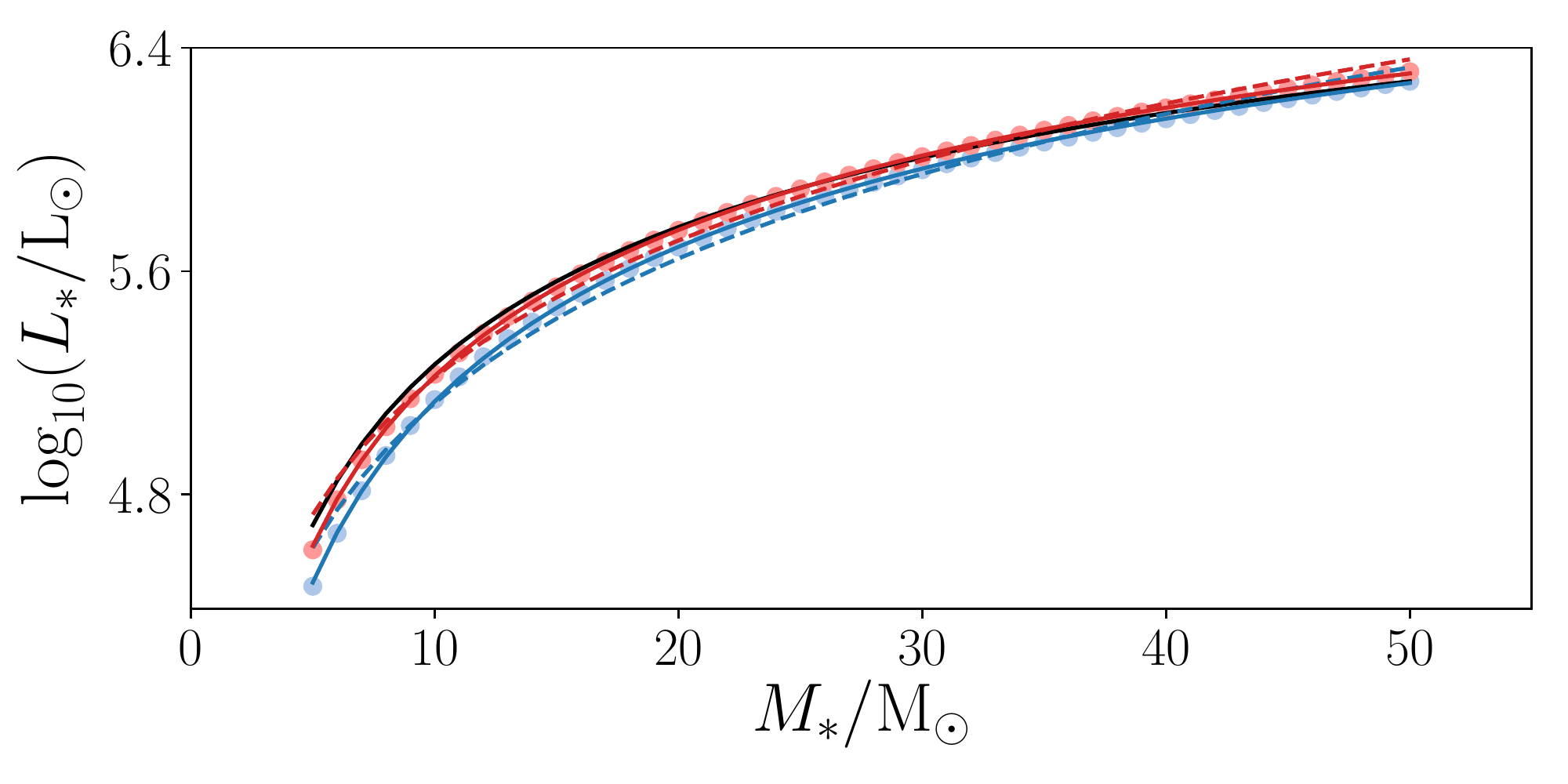}
    \includegraphics[width=\columnwidth]{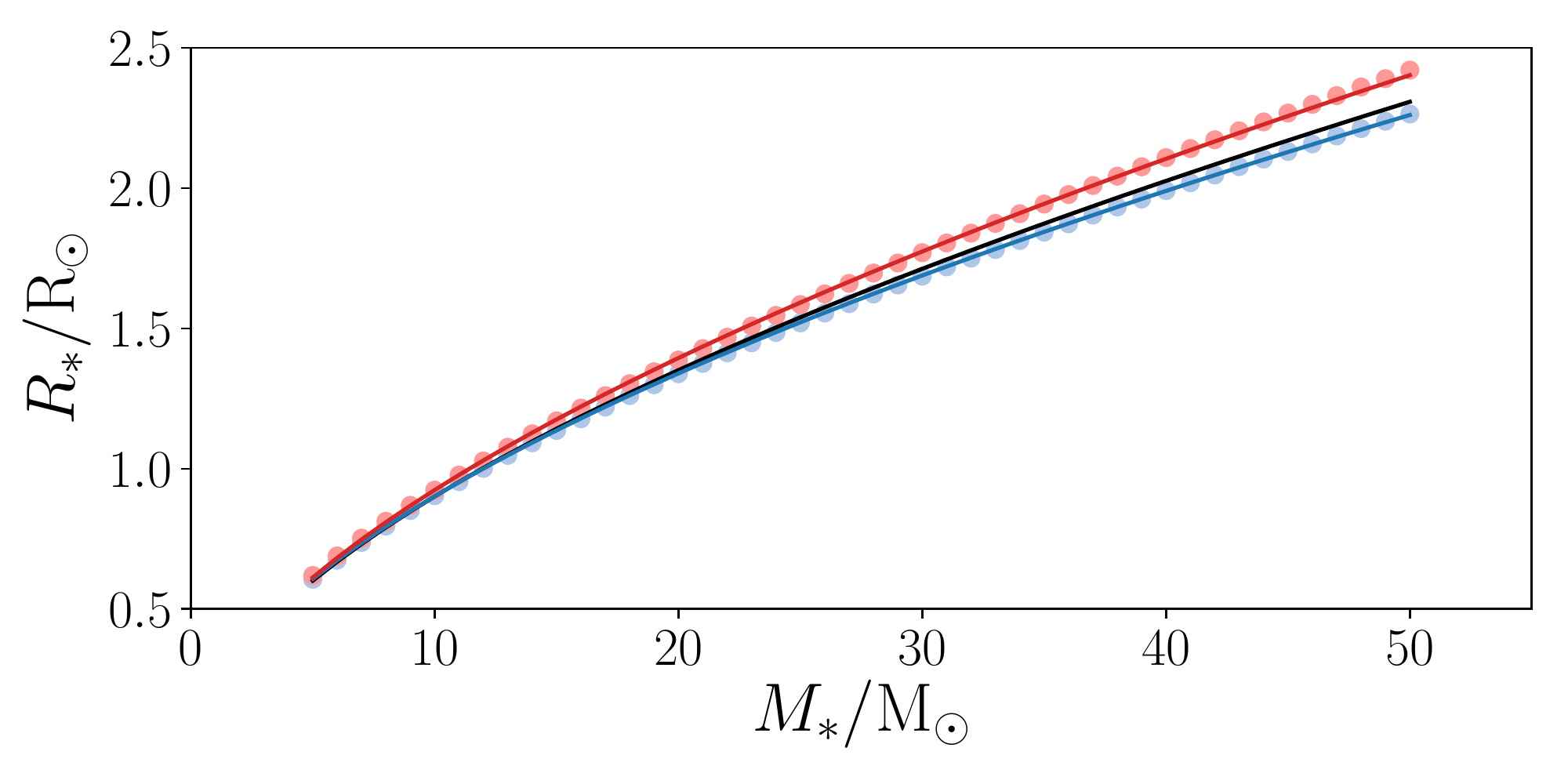}
\caption{Mass-luminosity (top panel) and mass-radius relations (bottom panel) for (1) pure helium (blue), and (2) mixture of He/C/O (red) stellar models with solar metallicity. Each point represents the conditions at $T=10^6$ K from a \texttt{MESA} model. Solid lines are Eq.\,(\ref{eq:empiricalmesa}). Dashed lines in panel a are Eq.\,(\ref{eq:empiricalmesa2}). Black lines are relations from \cite{1992A&A...263..129S}. }  
  \label{fig:mesa}
\end{figure}

I use the test suite model \texttt{make\_massive\_with\_ uniform\_composition} from \texttt{MESA-r9793} to construct models of pure helium and mixture of He/C/O. I assume these are representative of WNE and WC interiors, respectively. See \S\ref{sec:unstableenvelopesolutions} for the chemical mass fractions used. The evolution is stopped after the He core mass fraction is reduced by $2\%$. The luminosities and radii at $10^{5.6}$\,K are shown in Figure\,\ref{fig:mesa} for stellar masses between 5 and 50\,$\msun$. The fitted empirical relations are
\bea
\ell_{\rm WNE} &=& -0.62m^2 + 3.29m+2.46, \nonumber \\
\mathcal{R}_{\rm WNE} &=& 0.58m - 0.63, \nonumber \\
\ell_{\rm WC} &=& -0.47m^2 + 2.82m+2.87, \nonumber\\
\mathcal{R}_{\rm WC} &=& 0.59m - 0.63, \label{eq:empiricalmesa}
\eea
where $\ell\equiv{\rm log_{10}}(L_*/\lsun)$, $\mathcal{R}\equiv{\rm log_{10}}(R_*/\rsun)$, and $m\equiv{\rm log_{10}}(M_*/\msun)$. See Fig. \ref{fig:mesa} for a comparison with \cite{1992A&A...263..129S}. Linear fits to the luminosities are shown for convenience:
\bea
\ell_{\rm WNE} &=& 1.72m+3.41, \nonumber \\
\ell_{\rm WC} &=& 1.63m+3.59.  
\label{eq:empiricalmesa2}
\eea

An inlist is available at the end of Appendix\,B that has been verified for \texttt{MESA-r10108}. The mixing length in our $23\,\msun$ inlist is tuned to reproduce the G12 results.

\section*{Appendix\,B \\ Critical Mass Loss Relations}
\label{sec:appendixb}
The minimum mass loss rate, $\dot{m}\equiv\mathrm{log_{10}}(\dot{M}/(\msun\, \mathrm{yr}\inv))$, for an iron bump to harbour a sonic point in a hydrogen-free WR stellar model (Eq.\,\ref{eq:mdotmin}) is approximately:
\beq
\dot{m}_{\rm sp} = 
    \begin{cases}
      -11.6 - 1.6\ell + 2.6\ell_{\rm sp}(Z), & \mathrm{if} \ \ell \le \ell_{\rm sp}(Z)\\
      -11.6 + \ell, &  \mathrm{if} \ \ell \ge \ell_{\rm sp}(Z)
    \end{cases}
\eeq
with a global minimum at $\ell_{\rm sp}(Z)\equiv6.2-20Z$, where $0.01\le Z \le 0.03$ is the metallicity. 

An approximation for the WR mass loss relation $\mdot_{\alpha}$ (Eq.\,\ref{eq:mdotbubble}) follows a double logarithm of $L_*$,
\beq
\dot{m}_{\alpha} = \dot{m}_0 +  0.63\, \mathrm{log_{10}}\left(\frac{\ell-\ell_0}{12.4 - \ell_0 - \ell}\right),
\label{eq:ymdotbubble}
\eeq
where
\bea
\ell_b(Z) &\equiv& [6.00,\ 5.72,\ 5.55], \label{eq:lumb} \\
\ell_0(\alpha,Z)&\equiv& 0.44\, \mathrm{log_{10}}(\alpha) + \ell_b(Z),\\
\dot{m}_0(\alpha,Z) &\equiv& -0.68\mathrm{log_{10}}(\alpha)-[4.63,\ 4.53,\ 4.50],
\eea
for metallicities at $Z=[0.01, 0.02, 0.03]$ and $0.05\leq\alpha\le1$. Note that Eq.\,(\ref{eq:ymdotbubble}) is undefined for $\dot{m}_{\alpha}<\dot{m}_{\rm sp}$ since there is no sonic point. For $\alpha=1$, the maximum mass loss rate to harbour extended winds (i.e., inflated envelopes), $\mdot_b\equiv\mdot_{\alpha=1}$, asymptotically approaches zero at the minimum luminosity for envelope inflation in hydrostatic models $\ell_b$. The values for $\ell_b$ are fitted parameters from Eq.\,(\ref{eq:ymdotbubble}) and (\ref{eq:mdotbubble}) which overestimate $L_b$ by $\sim0.15$ dex.

\begin{lstlisting}
&star_job
    relax_initial_to_xaccrete = .true.
    num_steps_to_relax_bomposition = 50
    set_HELM_ion_neutral_blends = .true.
    max_logRho_neutral_HELM = -20    
    relax_to_this_tau_factor = 1d0
    dlogtau_factor = 0.1
    relax_initial_tau_factor = .true.
    set_initial_dt = .true.
    years_for_initial_dt = 1d-8
/
&controls
    max_number_backups = 150
    max_number_retries = 600
    initial_mass = 23d0
    Zbase = 0.02d0
    varcontrol_target = 1d-3
    dX_nuc_drop_limit = 2d-2 ! 5d-3
    delta_lgT_bntr_limit = 0.01  
    delta_lgT_bntr_hard_limit = 0.03
    delta_lgRho_bntr_limit = 0.05  
    delta_lgRho_bntr_hard_limit = 0.15
    cool_wind_RGB_scheme = ''
    cool_wind_AGB_scheme = ''
    accrete_same_as_surface = .false. 
    accrete_given_mass_fractions = .true. 
    ! Pure He and stopping condition
    num_accretion_species = 1
    accretion_species_id(1) = 'he4'
    accretion_species_xa(1) = 1d0
    xa_bentral_lower_limit_species(1)='he4'
    xa_bentral_lower_limit(1) = 98d-2
    ! MLT++ off:
    okay_to_reduce_gradT_excess = .false.  
    mixing_length_alpha = 0.325!tuned to G12
    mesh_Pgas_div_P_exponent = 0.3
    max_allowed_nz = 20000
/
\end{lstlisting}
\bibliographystyle{apj}

\end{document}